\documentstyle[12pt,titlepage,epsf]{article}  
\newcommand{\nin}{\noindent}
\newcommand{\be}{\begin{equation}}
\newcommand{\ee}{\end{equation}}
\newcommand{\bea}{\begin{eqnarray}}
\newcommand{\eea}{\end{eqnarray}}

\newcommand{\nonu}{\nonumber\\}

\newcommand{\ol}{\overline}

\begin{document}

\hfill   NTUA-92/00

\vspace{1cm}

\begin{center}
{\bf QED in a Strong External Magnetic Field:}

{\bf Beyond the Constant Mass Approximation}

\vspace{1cm}

{\bf J.Alexandre, K. Farakos, G. Koutsoumbas

Department of Physics, National Technical University of
Athens,

Zografou Campus, 157 80 Athens, GREECE}
\vspace{15mm}

Abstract
\end{center}

We solve the Schwinger-Dyson equations for QED in 2+1 or 3+1 dimensions 
in the presence of a strong homogeneous external magnetic field. 
The magnetic field is assumed strong enough, so that the lowest
Landau level approximation holds, but the usual
assumption of a momentum-independent self-energy is not made.  
In 2+1 dimensions, the scaling with logarithm changes to a square root 
dependence on the magnetic field, but
the most spectacular result takes place in 3+1 dimensions, where the
constant mass approximation turns out to be unreliable and the 
(momentum-dependent) dynamical mass is larger by several orders of 
magnitude compared to what has been found till now using the 
constant mass approximation.

\newpage

\section{Introduction}

Magnetic catalysis of chiral symmetry breaking  
is well known by now as a concrete example of 
dynamical symmetry breaking, both
in 2+1 and 3+1 dimensions. 
It is, of course, interesting from the field-theoretical point of view to 
have an example of dynamical symmetry breaking where explicit calculations
may be performed. 
The non-perturbative tool used
has been a consistent truncation of the SD equations based on the domination
of the dynamics by the lowest Landau level (LLL) in the regime of
strong magnetic fields.
The results have shown that the presence of
a homogeneous external magnetic field is enough to trigger dynamical mass
generation even at very weak attractive coupling between the fermions. 

The long history of the subject involves among others 
the very interesting issue of its role in models of the QCD vacuum 
\cite{savvidy} and a possible involvement in a new phase of QED with
spontaneous chiral symmetry breaking, which has been claimed to offer 
an explanation of the $e^+e^-$ pairs observed in the heavy ion experiments 
in GSI \cite{mal}.

The Standard Electroweak Model in the presence of a constant magnetic 
field has been examined by Ambjorn and Olesen \cite{ambjorn}. 
Their result is very 
briefly that, for strong enough magnetic field strength, they obtained
a $W$ and $Z$ condensate solution and a
lattice of 
vortex lines is formed, while at even stronger magnetic fields a phase 
transition to the symmetric phase takes place; the critical 
temperature needed for the symmetry restoration is substantially smaller 
than the one without magnetic field. Thus, it is seen that the magnetic
field may induce a transition from the broken to the symmetric phase of the 
Electroweak Standard Model. These ideas of symmetry restoration in the 
presence of a magnetic field 
are discussed also earlier (see \cite{linde}).  

The ground state of the electroweak theory has also been studied 
in the presence of a hypermagnetic field to the end of stregthening the 
electroweak phase transition so that it might give rise to baryogenesis
within the Standard Model \cite{shaposhnikov}. Moreover, the numerical 
work done and the contribution of ring diagrams in the effective potential
show that the hypermagnetic field strengthening of the phase transition 
is not enough to satisfy all the standard model baryogenesis conditions.

Models for the formation of the strong magnetic fields that
have been observed on the scale of galaxies are strongly based on the
existence of extremely intense primordial magnetic fields \cite{early}.
These astrophysical and cosmological considerations are another field 
where magnetic field induced chiral symmetry breaking is relevant.

The model of a QED field theory with fermions has been suggested to 
describe the long-wavelength limit of the high-$T_c$
superconducting cuprates \cite{shankar}. 
Aspects of the behaviour of these materials,
such as the scaling of their thermal conductivity, may find their
explanation in the framework of the model under the influence of an
external magnetic field \cite{farakmavro}.

The theory of dimensional reduction and magnetic catalysis has been 
developed by several groups \cite{gusynin99, mimi, reuter, shpagin}.
They have treated the problem of a constant
external magnetic field acting on either Nambu Jona Lasinio model 
or QED in 2+1 or 3+1
dimensions. It was found that the magnetic field is actually a strong 
catalyst and even the weakest possible attraction between the fermions 
is enough for dynamical mass generation. In addition, in 2+1 dimensions,
mass is generated dynamically for any number of flavours 
and not only for $N \le 4,$ as is the case 
without magnetic field \cite{pisarsky}
and the critical temperature for symmetry restoration is much smaller than 
$\sqrt{eB}.$

There have also been found solutions \cite{farakout} 
of the Schwinger-Dyson equations 
in the limit of weak magnetic field for (2+1)-dimensional QED with 
both quenched and dynamical fermions \cite{farakout}. Also, if there is a 
coupling of the fermions to scalar fields in the presence of a magnetic
field, a strong fermion condensate appears analogue to $\sqrt{eB}$ 
\cite{incera}.

All of the above approaches have used the constant mass approximation, 
that is the self energy was supposed to be independent from the momentum. 
In this paper we try to make a step towards discarding this rather crude
assumption and see the consequences in both the 2+1 and the 3+1 dimensional
theories. To be able to proceed, we restrict ourselves to the strong field
regime.

\section{Fermions in a constant magnetic field}

To fix our notations we shortly review here the characteristics of
fermions in a constant external magnetic field, both in 2+1 and 3+1
dimensions. 

The model we are going to consider is described by the Lagrangian density:

\be
{\cal L}=-\frac{1}{4}f_{\mu\nu}f^{\mu\nu}+i\ol\Psi D_\mu\gamma^\mu\Psi 
-m\ol\Psi\Psi,
\ee

\nin where $D_\mu = \partial_\mu+i g a_\mu + i e A^{ext}_\mu,$ 
$a_\mu$ is an abelian quantum gauge field, $f_{\mu \nu}$ is the
corresponding field strength, and $A^{ext}_\mu$ describes an 
external electromagnetic field; in this work it will be a 
homogeneous magnetic field, constant in time. 
Notice that the fermions feel both the quantum and the external 
gauge fields, however we have allowed for different coupling 
constants, g and e. In three dimensions it makes sense to keep 
different couplings \cite{shankar}, 
while in 3+1 dimensions g is set equal to e.
We recall the usual definition $g^2\equiv 4\pi\alpha$.

We will choose the ``symmetric" gauge

\be
A^{ext}_1(x)=-\frac{B}{2}x_2,~~A^{ext}_2(x)=+\frac{B}{2}x_1,
\ee

\nin with the remaining components of the potential 
vanishing both in 2+1 and 3+1 dimensions.

\nin We notice for further reference that this potential 
has the property:

\be\label{propA}
x^\mu A^{ext}_\mu(y)+y^\mu A^{ext}_\mu(x)=0
\ee

\nin We know from the work of Schwinger 
\cite{schwinger} that the fermion propagator is given by:

\be
S(x,y)=e^{iex^\mu A^{ext}_\mu(y)}\tilde S(x-y),
\label{phase}
\ee

\nin where the translational invariant propagator $\tilde S$ 
has the following Fourier transform:

\bea\label{schwingerrep}
\tilde S(p)&=&\int_0^\infty ds \exp is\left(p_\|^2-p_\bot^2
\frac{\tan(|eB|s)}{|eB|s}-m^2\right)\nonu
&\times&
\left[(p^\|\gamma^\|+m)(1+\gamma^1\gamma^2\tan(|eB|s))-p_\bot\gamma_\bot
(1+\tan^2(|eB|s))\right]
\eea

\nin where $p^\bot \equiv (p^1,p^2)$, $p^\| \equiv p^0$ in 2+1 dimensions and
$p^\| \equiv (p^0,p^3)$ in 3+1 dimensional space time.
Similar notations hold for the $\gamma$-matrices.\\
There exists an additional representation for the 
fermion propagator, namely the Landau level expansion \cite{chodos}.  
This expansion is particularly useful in the strong field regime,
$|eB|>>m^2$, since one may keep only the first term and get 
the so-called lowest Landau level approximation (LLL) 
\cite{gusynin99,lee}:

\be\label{freeprop}
\tilde S_{LLL}(p)=ie^{-p_\bot^2/|eB|}\frac{p^\|\gamma^\|+m}{p_\|^2-m^2}
\left(1-i\gamma^1\gamma^2sg(eB)\right)
\ee

\nin  It is this strong field regime that we will consider in this paper.

\section{Schwinger-Dyson equation with an external magnetic field}

The Schwinger-Dyson equation for the fermion propagator 
in $d$ dimensions is given by the expression:

\be\label{SD}
G(x,y)=S(x,y)-4\pi\alpha\int d^dz_1d^dz_2d^dz_3d^dz_4
S(x,z_1)\gamma^\mu G(z_1,z_2)\Gamma^\nu(z_2,z_3,z_4) 
G(z_4,y)\Delta_{\mu\nu}(z_3,z_1),
\ee

\nin where $G$ and $\Delta_{\mu\nu}$ are the 
full fermion and photon propagators and 
$\Gamma^\nu$ the full vertex. We show now perturbatively that $G$ 
factors in the same way as the bare propagator $S$ in equation (\ref{phase}), 
in other words it can be written in the form:

\be
G(x,y)=e^{iex^\mu A^{ext}_\mu(y)}\tilde G(x-y). \label{fphase}
\ee

\nin Making a loop expansion 
in (\ref{SD}) and keeping only the tree level and the one loop terms 
we can write:
\be\label{sdexp}
G(x,y)=S(x,y)-4\pi\alpha\int d^dz_1d^dz_2
S(x,z_1)\gamma^\mu S(z_1,z_2)\gamma^\nu
S(z_2,y)D_{\mu\nu}(z_2,z_1)+...
\ee
\nin Note that we have used the bare vertex  
$\Gamma^\nu(z_2,z_3,z_4)=\gamma^\nu\delta(z_2-z_3)\delta(z_3-z_4),$
and that to this order in the loop expansion only the free photon 
$D_{\mu\nu}$ and fermion $S$ propagators appear.

Let us now isolate the one-loop term in the
expansion (\ref{sdexp}). This term reads:

\bea
G^{(1)}(x,y)&=&-4 \pi \alpha \int d^dz_1d^dz_2S(x,z_1)
\gamma^\mu S(z_1,z_2)\gamma^\nu
S(z_2,y)D_{\mu\nu}(z_2,z_1)\nonu
&=&-4 \pi \alpha \int d^dz_1d^dz_2\tilde S(x-z_1)\gamma^\mu \tilde S(z_1-z_2)
\gamma^\nu \tilde S(z_2-y) D_{\mu\nu}(z_2-z_1)\nonu
&&~~~~~~~~\times \exp ie\left(x^\mu A_\mu^{ext}(z_1)+z^\mu_1 
A^{ext}_\mu(z_2)+z_2^\mu A^{ext}_\mu(y)\right)
\eea

\nin Using the change of variables $z_1=y-u$, $z_2=x+v$ and 
writing $r \equiv x-y$, we obtain:

\bea
G^{(1)}(x,y)&=&-4 \pi \alpha \int d^dud^dv\tilde S(r+u)
\gamma^\mu\tilde S(-r-u-v)
\gamma^\nu\tilde S(r+v) D_{\mu\nu}(r+u+v)\nonu
&&\times \exp ie\left(x^\mu A^{ext}_\mu(y-u)+
(y-u)^\mu A^{ext}_\mu(x+v)+(x+v)^\mu A^{ext}_\mu(y)\right)
\eea

\nin Taking advantage of the property (\ref{propA}) 
of the potential $A^{ext}_\mu$,
we finally obtain:

\be
G^{(1)}(x,y)=e^{iex^\mu A^{ext}_\mu(y)}\tilde G^{(1)}(r),
\ee

\nin where

\be
\tilde G^{(1)}(r)=-4 \pi \alpha \int d^dud^dve^{iev^\mu A^{ext}_\mu(u)}
\tilde S(r+u)\gamma^\mu\tilde S(-r-u-v)\gamma^\nu 
\tilde S(r+v) D_{\mu\nu}(r+u+v)
\ee

\nin which is exactly equation (\ref{fphase}). To be precise, the full
propagator equals a phase times a translational invariant quantity. 
Proceeding the same way
we can see that all the higher loop corrections behave as the one loop,
showing perturbatively that $G$ factorizes in the same way as $S$.

We can also see that the 
fermion self energy $\Sigma=G^{-1}-S^{-1}$ has also the
same structure as $S,$ since the Schwinger-Dyson equations can be written as:

\be\label{SD2}
\Sigma(x,y)=4\pi\alpha\gamma^\mu\int d^dz_1d^dz_2
G(x,z_1)\Gamma^\nu(y,z_1,z_2)
\Delta_{\mu\nu}(z_2,x),
\ee

\nin such that the one-loop self energy is:

\bea
\Sigma^{(1)}(x,y)&=&
4\pi\alpha\gamma^\mu G(x,y)\gamma^\nu D_{\mu\nu}(y,x)\nonu
&=&e^{iex^\mu A^{ext}_\mu(y)}\hat\Sigma(x-y)
\eea

\nin 
where $\hat\Sigma(r)=4\pi\alpha\gamma^\mu
\hat G(r)\gamma^\nu D_{\mu\nu}(r)$. The higher order corrections to
$\Sigma$ will have the same phase.

Since we will deal with the LLL approximation only, we cannot define the 
inverse of the fermion propagator (\ref{freeprop}) because of the 
projection operator $(1-i\gamma^1\gamma^2)/2;$ 
thus we cannot define the LLL approximation for $\Sigma(x,y).$
For this reason we will prefer to use the
Schwinger-Dyson equation (\ref{SD}) and not (\ref{SD2}). We will also 
take the bare vertex, so that
our starting equation is:

\be\label{SDprime}
G(x,y)=S(x,y)-4\pi\alpha\int d^dz_1d^dz_2S(x,z_1)
\gamma^\mu G(z_1,z_2)\gamma^\nu
G(z_2,y)\Delta_{\mu\nu}(z_2,z_1)
\ee

\nin We rewrite equation (\ref{SDprime}) in terms of the translational 
invariant fermion propagators $\tilde S(r)$ and $\tilde G(r)$ 

\be
\tilde G(r)=\tilde S(r)
-4\pi\alpha\int d^dz_1d^dz_2
\tilde S(r-z_1)\gamma^\mu
\tilde G(z_1-z_2)\gamma^\nu\tilde G(z_2)\Delta_{\mu\nu}(z_2-z_1)
e^{ie(r-z_2)^\mu A^{ext}_\mu(z_1)}
\ee

\nin where $r=x-y$.
Making a Fourier transform and performing the integration 
over $z_1$ we obtain:

\be\label{sd3}
\tilde G(k)=\tilde S(k)-4\pi\alpha \tilde I(k),
\ee

\nin where 

\be
\tilde I(k)=\int d^dz\int\frac{d^dp}{(2\pi)^d}\frac{d^dq}{(2\pi)^d}
\tilde S(p-q+eA^{ext}(z))\gamma^\mu\tilde G(p)\gamma^\nu
\tilde G(k-eA^{ext}(z))\Delta_{\mu\nu}(q)e^{iz(q-p+k)}.
\ee

\nin
We will look for a solution of (\ref{sd3})
where the translational part of the full propagator
reads in the spirit of the LLL approximation (equation (\ref{freeprop})):

\be\label{fullprop}
\tilde G_{LLL}(p)=ie^{-p_\bot^2/|eB|}\frac{Z_pp^\|\gamma^\|+M_p}
{Z_p^2p_\|^2-M_p^2}\left(1-i\gamma^1\gamma^2sg(eB)\right),
\ee

\nin where $Z_p$ is the wave function renormalization and $M_p$
the dynamically generated mass.

\section{Magnetic catalysis in 2+1 dimensions}

\subsection{Integral equations}

We will look for a solution of (\ref{sd3}) with the 
expressions (\ref{freeprop}) with $m=0$ and (\ref{fullprop})
substituted for the free and the full fermion propagator
respectively. We are using four-component representation for the fermions
and the $\gamma$ matrices. 
To obtain the equations that $Z_k$ and $M_k$ must satisfy,
we multiply $\tilde I(k)$ respectively by $(Z_kk^0\gamma^0-M_k)$ and
$\gamma^0(Z_kk^0\gamma^0-M_k)$. We do a Wick rotation
and then perform the integration over 
$z_0=iz_3$, which leads to a factor $2\pi\delta(q_3-p_3+k_3)$. 
We obtain then the equations:

\bea
\mbox{tr}\left\{\left(Z_kik_3\gamma^0-M_k\right)I(k)\right\}&=&
-\frac{16i}{k_3}\int d^2z_\bot\int\frac{d^3p_E}{(2\pi)^3}
\frac{d^2q_\bot}{(2\pi)^2}e^\phi\frac{Z_pp_3}{Z_pp_3^2+M_p^2}
{\cal D}(p_3-k_3,q_\bot)\nonu
\mbox{tr}\left\{\gamma^0\left(Z_kik_3\gamma^0-M_k\right)I(k)\right\}&=&
-\frac{16}{k_3}\int d^2z_\bot\int\frac{d^3p_E}{(2\pi)^3}
\frac{d^2q_\bot}{(2\pi)^2}e^\phi\frac{M_p}{Z_pp_3^2+M_p^2}
{\cal D}(p_3-k_3,q_\bot), \nonumber
\eea

\nin where 
$ \phi \equiv iz_\bot(q_\bot-p_\bot+k_\bot)-\frac{1}{|eB|}
\left[(p_\bot-q_\bot+eA^{ext}(z))^2+p_\bot^2+(k_\bot-eA^{ext}(z))^2\right]
$
and ${\cal D}(q_3,q_\bot)=-iD_{00}(q_E).$ We note that only the
component $D_{00}$ of the photon propagator plays a role in the 
Schwinger-Dyson equations, 
due to the projection operator $(1-i\gamma^1\gamma^2)/2$ which appears 
in the fermion propagators when the LLL approximation is used.\\
The integrations over $p_\bot$ and $z_\bot$ are straightforward.
We also have the auxiliary relations:
\bea
\mbox{tr}\left\{\left(Z_kik_3\gamma^0-M_k\right)
\left(\tilde G(k)-\tilde S(k)\right)\right\}&=&
4ie^{-k_\bot^2/|eB|}(1-Z_k)\nonu
\mbox{tr}\left\{\gamma^0\left(Z_kik_3\gamma^0-M_k\right)
\left(\tilde G(k)-\tilde S(k)\right)\right\}&=&
4e^{-k_\bot^2/|eB|}\frac{M_k}{k_3},\nonumber
\eea
\nin such that (\ref{sd3}) yields the final integral equations:

\bea\label{intequa3a}
\kappa(1-Z_\kappa)&=&
\frac{\tilde\alpha}{\pi}\int_{-\infty}^\infty dv \int_0^\infty du
e^{-u^2/2}\frac{uvZ_v}{Z^2_vv^2+\mu^2_v}{\cal D}(v-\kappa,u)
\eea
\bea\label{intequa3b}
\mu_\kappa&=&
\frac{\tilde\alpha}{\pi}\int_{-\infty}^\infty dv \int_0^\infty du
e^{-u^2/2}\frac{u\mu_v}{Z^2_vv^2+\mu^2_v}{\cal D}(v-\kappa,u)
\eea

\nin We have introduced the dimensionless variables
$\tilde\alpha \equiv \alpha/\sqrt{|eB|}$, $u \equiv \sqrt{q_\bot^2/|eB|}$,
$v \equiv p_3/\sqrt{|eB|}$, $\kappa \equiv k_3/\sqrt{|eB|}$ and
$\mu_\kappa \equiv M_k/\sqrt{|eB|}$. 
The reader can check that the integral equations
(\ref{intequa3a}, \ref{intequa3b}) are consistent
with the ones in \cite{shpagin}. We note that $\mu_\kappa$
and $Z_\kappa$ depend only on the dimensionless ratio $\tilde\alpha$.

The free photon propagator $\Delta_{\mu \nu}$ will be used,  
since we show in the appendix that it receives no corrections in this case. 
In particular, it is shown that the one-loop polarization tensor
is suppressed by $\alpha|eB|^{-1/2}$, such that
in the LLL approximation we will use in 
(\ref{intequa3a}, \ref{intequa3b}) the Euclidean photon propagator

\be
{\cal D}(v-\kappa,u)=\frac{1}{u^2+(v-\kappa)^2}
\left(1-\frac{(v-\kappa)^2}{u^2+(v-\kappa)^2}\right)
+\lambda\frac{(v-\kappa)^2}{\left(u^2+(v-\kappa)^2\right)^2}
\ee

\nin where $\lambda$ is the gauge fixing parameter.

\subsection{Constant mass approximation}

Before proceeding with a detailed analysis of (\ref{intequa3a}, 
\ref{intequa3b}), we start with a first approximation, which consists 
in setting $Z_v=1$ and 
a constant dynamical mass $\mu_c$ in the integral equations, so that
in the Feynman gauge $\lambda=1$ equation (\ref{intequa3b}) reads:
\be
\mu^{const}_\kappa=\frac{\tilde\alpha}{\pi}\int_{-\infty}^\infty dv 
\int_0^\infty du e^{-u^2/2}
\frac{u\mu_c}{v^2+\mu_c^2}\frac{1}{u^2+(v-\kappa)^2}
\ee

\nin The integration over $v$ is done by the residue theorem and 
leads to:
\be\label{constmassx}
\mu_\kappa^{const}=\tilde\alpha\int_0^\infty du e^{-u^2/2}\frac{u+\mu_c}
{\kappa^2+(u+\mu_c)^2}
\ee
\nin Thus the constant dynamical mass is obtained by solving the
equation $\mu_{\kappa=0}^{const}=\mu_c$ which reads:
\be\label{constgap3}
\mu_c=\tilde\alpha\int_0^\infty du e^{-u^2/2}\frac{1}{u+\mu_c}
\ee
\nin The study of (\ref{constgap3}) has also been done in \cite{farakmavro}.\\ 
Putting $Z_v=1$ as a first approximation and taking a constant mass 
in equation (\ref{intequa3b})
leads to the following second approximation to $Z_\kappa:$
\be
Z_\kappa^{const}=1-
\tilde\alpha\int_0^\infty du e^{-u^2/2}\frac{1}{\kappa^2+(u+\mu_c)^2},
\ee
\nin such that its value for zero momentum is given by:

\be
Z_c=1-\tilde\alpha\int_0^\infty du e^{-u^2/2}\frac{1}{(u+\mu_c)^2}.
\ee

\nin We plot the dimensionless 
dynamical mass gap $\tilde\alpha^{-1}\mu_c/Z_c$  
versus $\sqrt{|eB|}/\alpha$ in figure \ref{dmb3} and compare with the 
mass gap obtained with $\mu_{\kappa=0}$ and $Z_{\kappa=0}$ solutions of
(\ref{intequa3a},\ref{intequa3b}) that we find in the next section.
We can see that the constant mass approximation leads to an over-evaluation
of the dynamical mass gap and that it gets worse as
the magnetic field increases.

\subsection{Momentum dependent solutions}

We now concentrate on the momentum dependence of the dynamical 
mass and the wave function renormalization to obtain a more precise 
analysis of the solution of (\ref{intequa3a},\ref{intequa3b}).\\
Let us first study the wave function renormalization 
$Z_\kappa$. We will actually consider the one-loop approximation,
consistently with our treatment of the Schwinger-Dyson equation
(\ref{SD}), where we considered the bare vertex. 
Since there is no pertubative solution of the 
integral equations for the dynamical mass $\mu_\kappa$,  
our numerical solution of the system of integral equations will 
correspond to some resummation of diagrams. 
The solution will be found through an iterative procedure: 
a trial function will be put 
in equations (\ref{intequa3a}, \ref{intequa3b}) to yield a first estimate of
$Z_\kappa$ and $\mu_\kappa.$ We will then take this output and 
substitute it back in the right hand sides of the integral equations 
to obtain better approximations. As we will see, this iterative procedure 
does finally converge to a solution.\\
The trial function will be given by the solution of a differential 
equation verified by $\mu_\kappa$ that will be derived now. This differential
equation is derived also in \cite{gusynin99} for the (3+1)-dimensional
model in what they called linearized approximation. \\

Let us now proceed with equation (\ref{intequa3b}).
We start by splitting the integration over $v$ in (\ref{intequa3b}) in
two parts:

\bea\label{split}
\mu_\kappa&\simeq&\frac{\tilde\alpha}{\pi}\int_{|v|<\kappa} dv \int_0^\infty du
e^{-u^2/2}\frac{u\mu_v}{Z^2_vv^2+\mu^2_v}{\cal D}(\kappa,u)\nonu
&+&\frac{\tilde\alpha}{\pi}\int_{|v|>\kappa} dv \int_0^\infty du
e^{-u^2/2}\frac{u\mu_v}{Z^2_vv^2+\mu^2_v}{\cal D}(v,u)
\eea

\nin Notice the approximations used:
${\cal D}(v-\kappa,u) \approx {\cal D}(\kappa,u)$ for $|v|<\kappa$ and
${\cal D}(v-\kappa,u) \approx {\cal D}(v,u)$ for $|v|>\kappa.$
Then the derivative with respect to $\kappa$ 
leads to the following equality:

\be
\mu_\kappa'\equiv\frac{d \mu_\kappa}{d \kappa} = 
\frac{2\tilde\alpha}{\pi}f_\kappa\int_0^\kappa dv 
\frac{\mu_v}{Z^2_vv^2+\mu^2_v}
\ee

\nin where we defined:

\be
f_\kappa \equiv \frac{d}{d\kappa}\left\{
\int_0^\infty du e^{-u^2/2}u{\cal D}(\kappa,u)\right\}
\ee

\nin A new derivative with respect to $\kappa$ finally gives:

\be\label{equadiff3}
f_\kappa\mu_\kappa''-f_\kappa'\mu_\kappa'
-\frac{2\tilde\alpha}{\pi}f_\kappa^2\frac{\mu_\kappa}
{Z_\kappa^2\kappa^2+\mu_\kappa^2}=0
\ee

\begin{figure}
\epsfxsize=10cm
\epsfysize=8cm
\centerline{\epsfbox{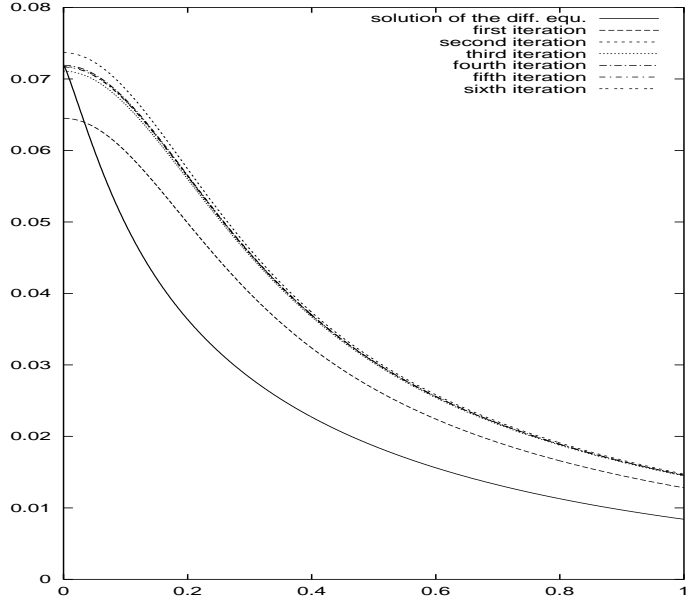}}
\caption{$\mu_\kappa$ versus $\kappa$ for
$\tilde\alpha=.03$ (d=2+1)}
\label{dmconv}
\end{figure}

\nin From now on we take $Z_\kappa=1$ in (\ref{equadiff3}). 
We should stress that this differential equation is only 
approximate, and the approximation (\ref{split}) is not
controled by any small parameter. 
Since we have no full control on the assumptions made, we consider 
the solution of (\ref{equadiff3}) 
only as a trial function which could help us with 
the solution of the integral equations.
It turns out that the iterative procedure described above converges 
quickly yielding a solution for $Z_\kappa$ and $\mu_\kappa$ 
reproducing itself after a small number of steps.
This may be seen in figure \ref{dmconv} for $\mu_\kappa$. We find
that the convergence of $Z_\kappa$ is even quicker. 
It turns out that the solution found as the limit
of the iterative procedure does not depend on the trial function
$\mu_\kappa$ that one starts with; however, taking
the solution of the differential equation gives a much better
convergence, that is the speed of convergence is increased.
As can be seen on figure \ref{dmconv}, $\mu_\kappa$ is almost flat
for $\kappa<\mu_{\kappa=0}$, as was also
emphasized in \cite{gusynin99} for 3+1 dimensions.

To solve (\ref{equadiff3}), we needed the initial conditions
that the function $\mu_\kappa$ has to satisfy. 
Suppose that $\mu_{\kappa=0}=\mu_0.$ In the Feynman gauge, the 
function $f_\kappa$ has the following behaviour around $\kappa=0$
\be
f_\kappa\sim-\frac{1}{\kappa}~~~~\mbox{as}~\kappa\to 0
\ee
\nin Thus if we wish $\mu_\kappa''$ to be finite in $\kappa=0,$
(\ref{equadiff3}) implies that:
\be
\mu_{\kappa=0}'=-\frac{2\tilde\alpha}{\pi\mu_0}.
\ee
\nin Next we have to determine $\mu_0$. We invoke the fact that  
$\mu_\kappa\to 0$ as $\kappa\to\infty$
and thus choose $\mu_0$ such that $\lim_{\kappa\to\infty}\mu_\kappa=0$.
As a side remark we note here that the trial function we obtain is not 
an even function of $\kappa;$ this is a consequence of the 
approximations made to get the differential equation (\ref{equadiff3}). 
The even function $\mu_\kappa$ is
found after plugging the trial function into the integral equations
(\ref{intequa3a}, \ref{intequa3b}). Equation (\ref{equadiff3}) has
been solved numerically through the use of a fourth order 
Runge-Kutta method. The estimate
of the function $f_\kappa$ as well as its derivative has been done 
using a Gauss-Hermite quadrature of order 40.

\begin{figure}
\epsfxsize=10cm
\epsfysize=8cm
\centerline{\epsfbox{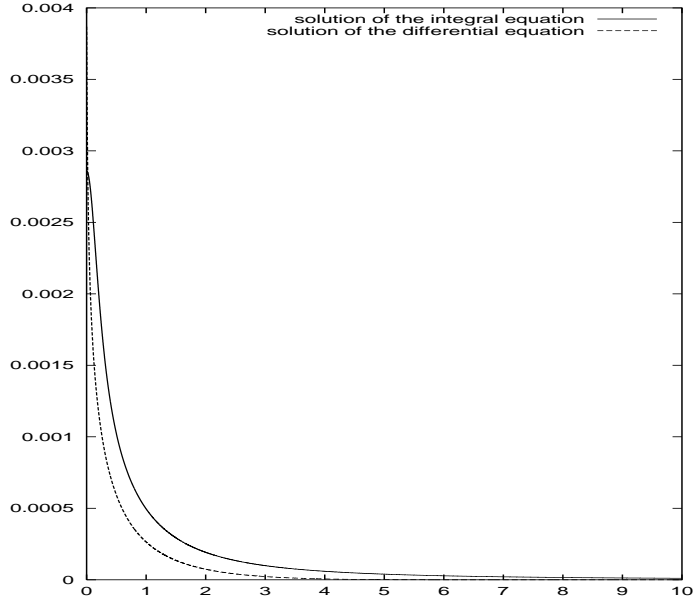}}
\caption{$\mu_\kappa/Z_\kappa$ versus $\kappa$ for 
$\tilde\alpha=.001$ (d=2+1)}
\label{dmm3}
\end{figure}

In figure \ref{dmm3} we plot 
the dimensionless mass gap $\mu_\kappa/Z_\kappa$ 
as a function of the dimensionless
momentum $\kappa$. One of the curves comes from the solution of 
the integral equations (using the iterative procedure), 
while for the sake of comparison we
also plot the solution of the differential equation (\ref{equadiff3}).
We note that in the differential equation we have set $Z_\kappa = 1;$
on the contrary, $Z_\kappa$ changes during the iterative procedure for the
solution of the integral equations. We observe that the quantity 
$\mu_\kappa/Z_\kappa$  
derived from the differential equation is very close to its final value
that is found after substituting it the integral equations. This explains
why the iterative procedure converges fast if we use $\mu_\kappa$ as input.
It is obvious from the figure that the mass gap decreases rather 
rapidly with increasing momentum and takes its maximum value for $\kappa=0.$
Thus it is to be expected that we will have big differences from the
constant mass approximation, which assumes the same value for  
$0 \le \kappa < \infty.$ The small momenta will be given the major role, 
in accordance with the fact that mass generation is an infrared phenomenon.

\begin{figure}
\epsfxsize=10cm
\epsfysize=8cm
\centerline{\epsfbox{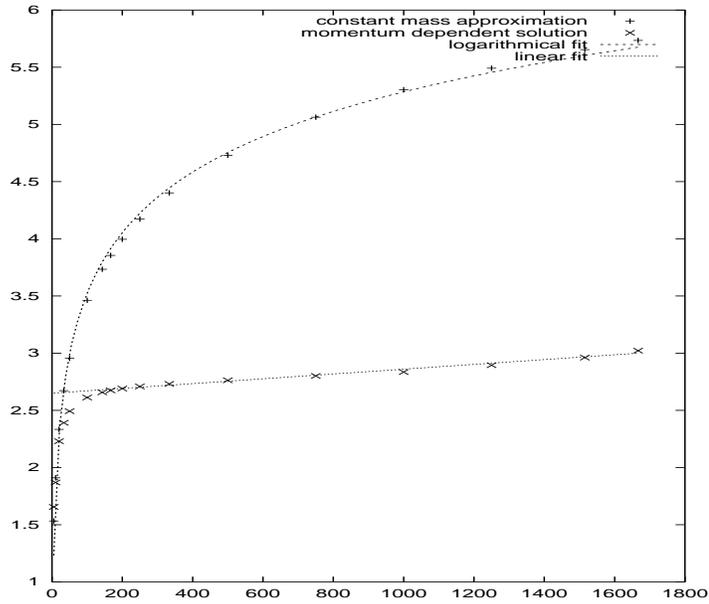}}
\caption{Dimensionless mass gap $\mu/\tilde\alpha$ versus 
$\sqrt{|eB|}/\alpha$ (d=2+1) }
\label{dmb3}
\end{figure}

In figure \ref{dmb3} we try to get contact with \cite{farakmavro} and
plot the dimensionless dynamical mass
$\mu/\tilde\alpha=M/\alpha$ versus $\sqrt{|eB|}/\alpha$ 
for two cases: the constant mass approximation obtained from 
(\ref{constgap3}) and the momentum dependent version, coming from the
solutions of the integral equations (\ref{intequa3a},\ref{intequa3b}).

In the constant mass case, we find a logarithmic fit, which reads: 

\be
\tilde\alpha^{-1}\mu^{fit}_{const}=m_0\ln\frac{\sqrt{|eB|}}{\alpha}
~~~~~~\mbox{with}~~~~~~m_0\simeq.765\pm.001
\ee

\nin or

\be
\mu^{fit}_{const}=
m_0\frac{\alpha}{\sqrt{|eB|}}\ln\frac{\sqrt{|eB|}}{\alpha}
\ee

\nin in agreement with the result of \cite{farakmavro}.

For the momentum dependent solution, we choose to plot the value of this
dynamical mass at $\kappa=0.$ versus $|e B|.$ 
The fit now is dramatically modified and reads: 

\be
\tilde\alpha^{-1}\mu^{fit}_{\kappa=0}=m_1+m_2
\frac{\sqrt{|eB|}}{\alpha}~~~~\mbox{with}
~~~~m_1=2.65\pm .01~~\mbox{and}~~m_2=(2.1 \pm 0.1) \times10^{-4} 
\label{exper}
\ee

\nin Of course, changing a logarithm to a square 
root is a fundamental difference   
from the previous case. We also observe the quantitative change in the 
value of the mass in comparison to the constant mass approximation: for the 
magnetic fields which are plotted in figure \ref{dmb3}, 
the mass is smaller by a 
factor of two to three. We should stress that the above results have been
produced using the lowest Landau level approximation, which holds provided 
the magnetic field is strong. This means that our results, shown in figure 
\ref{dmb3}, should not be trusted in the region near the origin of the 
horizontal axis. In this respect we observe in the figure that the 
data for the momentum dependent mass are well represented by the above fit
in the region of large magnetic fields down to the value 
$\sqrt{|eB|}/\alpha\simeq 100.$ We may interprete this change 
of behaviour of the data for small magnetic fields as been due to the 
LLL approximation, which is poor in this region; this may yield an estimate 
of the limit of its validity. We note here that with zero magnetic field,
the dynamical mass obtained by the Schwinger-Dyson equation in the 
constant mass approximation is bigger than the mass obtained solving 
the equations using a momentum dependent self energy for the fermion
\cite{pisarsky}.

\subsection{Fermion condensate}

The fermion condensate is given by

\be
\left<0|\ol\psi\psi|0\right>=
\lim_{y\to x}\mbox{tr} G(x,y)=
\mbox{tr}\int\frac{d^3p}{(2\pi)^3}\tilde G(p),
\ee

\nin which, in the LLL approximation, leads to:

\bea \label{running}
\left<0|\ol\psi\psi|0\right>&=&-4i\int\frac{d^2p_\bot}{(2\pi)^2}
e^{-p_\bot^2/|eB|}\int\frac{idp_3}{2\pi}\frac{M_p}{Z_p^2p_3^2+M_p^2}\nonu
&=&\frac{|eB|}{2\pi^2}
\int_{-\infty}^\infty dv\frac{\mu_v}{Z_v^2v^2+\mu_v^2}.
\eea

\nin In the constant mass approximation with $Z_v=1$ 
the integral (\ref{running}) yields the simple result:

\be\label{constcond}
\left<0|\ol\psi\psi|0\right>=\frac{|eB|}{2\pi}.
\ee
 
\begin{figure}
\epsfxsize=10cm
\epsfysize=8cm
\centerline{\epsfbox{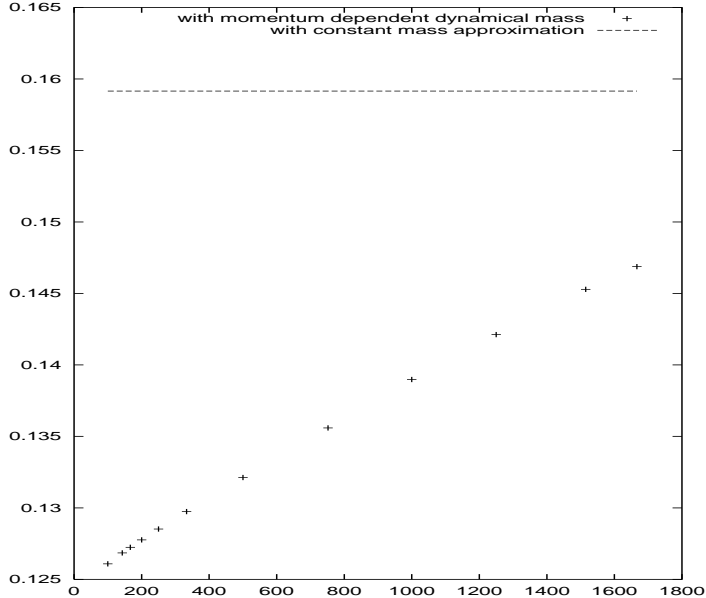}}
\caption{Dimensionless fermion condensate $<0|\ol\psi\psi|0>/|eB|$ 
versus $\sqrt{|eB|}/\alpha$ (d=2+1)}
\label{cond3}
\end{figure}

\nin It is worth noting that in this approximation the condensate does not 
depend on the dynamical mass $\mu_v.$
The condensate in the constant mass approximation grows  
linearly with $|eB|$ with a slope equal to $1/(2\pi)\simeq 0.1592$.  
We compared (\ref{constcond}) 
with the condensate given by equation (\ref{running}) using the 
momentum dependent functions $\mu_v$ and $Z_v$ found numerically in the
previous subsection. We show in figure 
\ref{cond3} the dimensionless condensate 
$\left<0|\ol\psi\psi|0\right>/|eB|$ versus $\sqrt{|eB|}/\alpha,$ 
both for the momentum dependent and the constant mass solutions. 
We see that the condensate varies linearly over a wide range
of $\sqrt{|eB|}/\alpha.$
We remark that the fermion condensate does not 
vanish in the limit $|eB|\to 0$
in 2+1 dimensions \cite{pisarsky}, but we cannot explore this limit 
within the LLL approximation, which is only reliable for strong 
magnetic fields.

\section{Magnetic catalysis in dimension d=3+1}

\subsection{Integral equations}

The techniques used in the (3+1)-dimensional model are similar to the ones 
developed in the previous section, so we will only  give the
results. \\
Starting again from the Schwinger-Dyson equations (\ref{SD}), 
we obtain in the same 
way as before

\bea\label{intequa4}
\kappa^2(1-Z_\kappa)&=&
\frac{\alpha}{\pi^2}\int d^2v\int_0^\infty du e^{-u^2/2}
\frac{uZ_v \vec\kappa.\vec v}{Z_v^2v^2+\mu_v^2}
{\cal D}(\vec v-\vec\kappa,u)\nonu
\mu_\kappa&=&\frac{\alpha}{\pi^2}\int d^2v\int_0^\infty du e^{-u^2/2}
\frac{u\mu_v}{Z_v^2v^2+\mu_v^2}{\cal D}(\vec v-\vec\kappa,u)
\eea

\nin where now $\vec\kappa$ is the two dimensional vector
representing two components of the dimensionless Euclidean momentum:  

\be
\vec\kappa=\frac{1}{\sqrt{|eB|}}(k_4,k_3)=\frac{1}{\sqrt{|eB|}}(-ik_0,k_3)
\ee

\nin In 3+1 dimensions, the corrections to the photon propagator are
not suppressed by inverse powers of the magnetic field, so we 
use the expression \cite{gusynin99}:

\be
D_{\mu\nu}(q)=-i\frac{g^\|_{\mu\nu}}{q^2+q_\|^2\Pi(q_\bot^2,q_\|^2)}
-i\frac{g^\bot_{\mu\nu}}{q^2}+i\frac{q^\bot_\mu q^\bot_\nu+
q^\bot_\mu q^\|_\nu+q^\|_\mu q^\bot_\nu}{q^4},
\ee

\nin where $\Pi(q_\bot^2,q_\|^2)$ is the polarization tensor and
$g_{\mu\nu}^\|$ and $g_{\mu\nu}^\bot$ represent the
restrictions of the metric tensor to the $(3,4)$ and $(1,2)$ 
directions respectively.
Because of the projection operator $(1-i\gamma^1\gamma^2sg(eB))/2$,
appearing in the fermion propagator in the LLL approximation, 
only the part of $D_{\mu\nu}$ containing $g_{\mu\nu}^\|$ will 
contribute to the Schwinger-Dyson equations. Therefore we will only have to 
consider the expression

\be\label{photprop4}
D^{LLL}_{\mu\nu}(q)=
-i\frac{g^\|_{\mu\nu}}{q^2+q_\|^2\Pi(q_\bot^2,q_\|^2)},
\ee

\nin where we will use for the polarization tensor the form \cite{reuter}:

\be\label{approxpi}
q_\|^2 \Pi(q_\|^2,q_\bot^2)\simeq-\frac{2\alpha |eB| }{\pi}
e^{-q_\bot^2/(2|eB|)},
\ee

\nin $\alpha$ being the QED coupling renormalized at the scale 
$\sqrt{|eB|}$.
Actually the approximation (\ref{approxpi}) is valid only for $q_\|^2>>M^2$
where $M$ is the dynamical mass of the fermions, but the latter will 
be very small compared to the typical momenta we will consider, so the
approximation is good. Therefore the photon propagator in dimension 3+1
will be:

\be\label{phot4}
D_{\mu\nu}^{LLL}(q)
=-i\frac{g^\|_{\mu\nu}}{q^2-\frac{2\alpha}{\pi}|eB|e^{-q_\bot^2/(2|eB|)}},
\ee

\nin such that the quantity ${\cal D}(\vec v-\vec\kappa,u)$ in (\ref{intequa4})
will be replaced by:

\be
{\cal D}(\vec v-\vec\kappa,u)=\frac{1}{u^2+(\vec v-\vec\kappa)^2+
\frac{2\alpha}{\pi}e^{-u^2/2}}
\ee

\nin The integral equation for $\mu_\kappa$ is consistent with the results of
\cite{gusynin99} where the authors do not consider the contribution 
of the wave function renormalization. It is known that in the dynamical 
symmetry breaking in QED, without external field, $Z_\kappa$ is equal 
to one, if we employ the 
Landau gauge \cite[sect.8.4]{miransky}. However, with an external 
magnetic field, the photon propagator (\ref{phot4}) is such that
we find $Z_\kappa\ne 1$ as can be seen in figure \ref{dz4}.

\subsection{Constant mass approximation}

We first have a look at the constant 
mass approximation of (\ref{intequa4}), 
to facilitate comparison with \cite{gusynin99}; we do not consider 
in this subsection the corrections to the photon propagator.
Denoting $\rho \equiv \vec \kappa^2$, we consider the integral 
equation for $\rho=0$ and $Z_v=1$ and set $\mu_\rho=\mu_v =\mu_c.$
The integration over $v$ then leads to the following gap equation:

\be\label{constgap4}
1=\frac{\alpha}{\pi}\int_0^\infty due^{-u^2/2}\frac{u}{\mu_c^2-u^2}
\ln\left(\frac{\mu_c^2}{u^2}\right)
\ee

\begin{figure}
\epsfxsize=10cm
\epsfysize=8cm
\centerline{\epsfbox{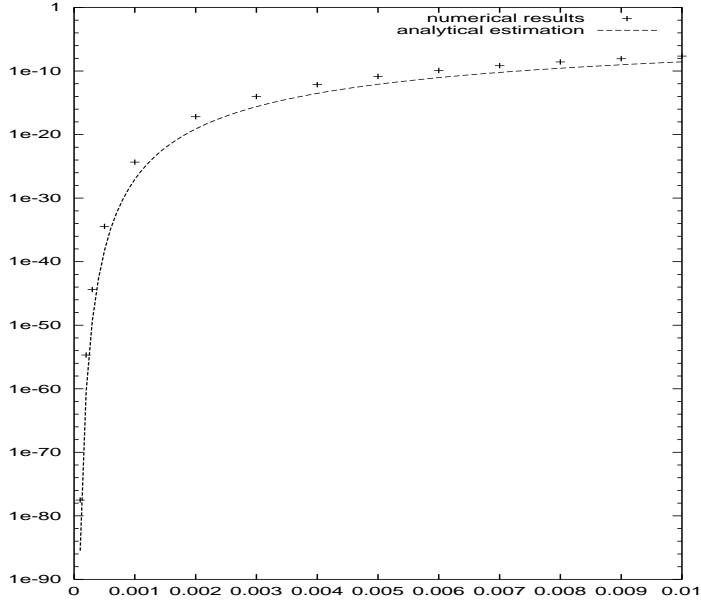}}
\caption{Dimensionless dynamical mass $\mu_c$ 
versus $\alpha$ in the constant mass approximation (d=3+1)}
\label{constdm4}
\end{figure}

\nin We plot in figure \ref{constdm4} the dynamical mass obtained
from (\ref{constgap4}) as well as the analytical estimate of 
\cite{gusynin99}, which is:

\be
\mu_c^{analyt}=C\exp\left[-\frac{\pi}{2}\left(
\frac{\pi}{2\alpha}\right)^{1/2}\right]  
\ee

\nin and choose $C=1$. We can see that this estimate agrees very well 
with our results. But we will see in the next 
paragraph that the momentum dependent dynamical mass gives completely 
different results from the constant mass approximation in 3+1 dimensions,
which was not the case in 2+1 dimensions. From the numerical point 
of view, this comes from the fact that
in 3+1 dimensions the dimensionless dynamical mass $\mu_c$ 
is not present in the left-hand side of the gap equation (\ref{constgap4}), 
contrary to what happened in (\ref{constgap3}), the corresponding
equation in 2+1 dimensions. 
The result is that the order of magnitude of the integral should be
$1/\alpha$ in 3+1 dimensions. This 
pushes the integral to very big values, that can be obtained by
extremely small values for $\mu_c.$ On the other hand, such problems are 
not present for the momentum dependent solution; this leads  
to a serious discrepancy, by several orders
of magnitude, between the momentum dependent solution of
(\ref{intequa4}) and the constant mass approximation which therefore
is not reliable in 3+1 dimensions. We note that this fact is independent
of our taking into account $Z_\kappa$, since the latter is very close to 1
and thus cannot give a change of several orders of magnitude
(see figure \ref{dz4}).
We also note that such problems do not arise in 2+1 dimensions, since, by
equation (\ref{constgap3}), the order of magnitude of the integral is
$\mu_c/\tilde \alpha$ in this case, so the smallness of $\tilde\alpha$
is compensated by a small number, such as $\mu_c$; thus moving to a
momentum dependent solution in 2+1 dimensions brings in some 
quantitative differences but not the huge ones characterising the
3+1 dimensions. It seems that the constant mass approximation 
in 2+1 dimensions is relatively reliable.

\subsection{Momentum dependent solutions}

We are looking for solutions of (\ref{intequa4}) which are 
even functions of the momentum. Thus we consider 
$\mu$ and $Z$ as functions of the variable $\rho \equiv \kappa^2=k^2/|eB|$.
The differential equation similar to (\ref{equadiff3}) is then obtained
in the same way and reads:

\be\label{equadiff4}
f_\rho\mu_\rho''-f_\rho'\mu_\rho'
-\frac{\alpha}{\pi}f_\rho^2\frac{\mu_\rho}{Z_\rho^2\rho+\mu_\rho^2}=0
\ee

\nin where the function $f_\rho$ is given by

\be
f_\rho=\frac{d}{d\rho}\left\{\int_0^\infty du e^{-u^2/2}u
{\cal D}(\sqrt\rho,u)\right\}.
\ee

\nin Equation (\ref{equadiff4}) is consistent with the one 
given in \cite{gusynin99} and we will set $Z_\rho=1$ to solve it.\\
Coming back to the integral equations (\ref{intequa4}), we can perform 
the $\vec v$-angular integration if we look for functions $\mu_\rho$ and
$Z_\rho$ depending on $\rho$ only. The result reads:

\bea\label{intequa4bis}
Z_\rho^{(1)}&=&1+
\frac{\alpha}{2\pi\rho}\int_0^\infty dv\int_0^\infty du e^{-u^2/2}
\frac{u}{v+\mu_v^2}\left[1-\frac{u^2+\rho+v+2\alpha/\pi\exp(-u^2/2)}
{\sqrt{\left(u^2+\rho+v+2\alpha/\pi\exp(-u^2/2)\right)^2-4\rho v}}\right]\nonu
\mu_\rho&=&\frac{\alpha}{\pi}\int_0^\infty dv\int_0^\infty du e^{-u^2/2}
\frac{u\mu_v}{Z_v^2v+\mu_v^2}\frac{1}
{\sqrt{\left(u^2+\rho+v+2\alpha/\pi\exp(-u^2/2)\right)^2-4\rho v}},
\eea

\nin where we have used the one-loop approximation $Z_\rho^{(1)}$
for the wave function renormalization $Z_v.$

\begin{figure}
\epsfxsize=10cm
\epsfysize=8cm
\centerline{\epsfbox{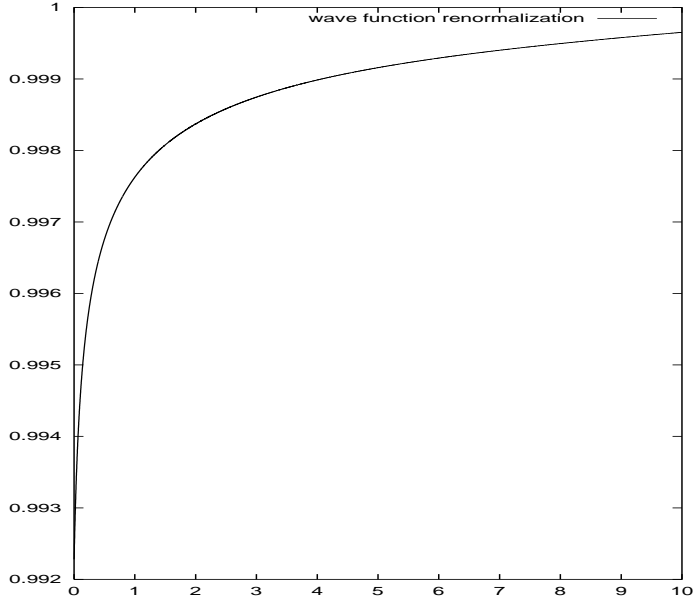}}
\caption{Wave function renormalization $Z_\rho$ 
versus $\rho$ for $\alpha=.01$ (d=3+1)}
\label{dz4}
\end{figure}

We solve numerically (\ref{intequa4bis}) in the same way as we solved
the equivalent set of equations in dimension 2+1. We first
solve the differential equation (\ref{equadiff4}) and use the 
solution as a trial function for the integral equations (\ref{intequa4bis}).
Then the outcome is used as feedback in the integral equations (\ref{dz4}) 
and the 
resulting iterative procedure is converging, somehow slower as compared to 
2+1 dimensions.
To find the initial conditions of $\mu_\rho$ satisfying (\ref{equadiff4}), 
we note that: 

\be
f_\rho\sim-\frac{1}{2\rho}~~~~\mbox{when}~\rho\to 0,
\ee

\nin so that (\ref{equadiff4}) implies that:

\be 
\mu_{\rho=0}'=-\frac{\alpha}{2\pi\mu_0}~~~~
\mbox{with}~~~\mu_0 \equiv \mu_{\rho = 0}, 
\ee

\begin{figure}
\epsfxsize=10cm
\epsfysize=8cm
\centerline{\epsfbox{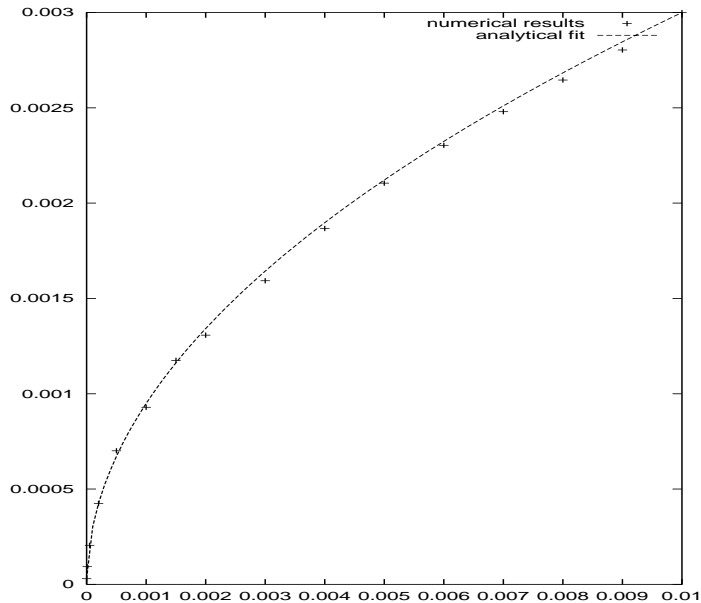}}
\caption{Dimensionless dynamical mass $\mu_{\rho=0}$
versus $\alpha$ (d=3+1)}
\label{dmc4}
\end{figure}

\nin if we demand that $\mu_\rho$ is twice differentiable as $\rho\to 0$. 
We then find $\mu_0$ by imposing 
the condition $\lim_{\rho\to\infty}\mu_\rho=0$, 
as we did in 2+1 dimensions.\\
In figure \ref{dmc4} we depict the evolution of the dimensionless
dynamical mass  at zero momentum, $\mu_{\kappa=0},$ with the 
coupling $\alpha$, as well as a fit which is given by: 

\be
\mu_{\rho=0}^{fit}=\mu_0\sqrt\alpha~~~~~\mbox{with}~~~~~\mu_0=.0300\pm 0.0001
\ee

\nin The dimensionless mass gap $(\mu_\rho/Z_\rho)_{\kappa=0}$ 
follows the same fit, since $Z_\rho\simeq 1$.
We may compare figures \ref{constdm4} and \ref{dmc4} and see the huge 
difference (by some orders of magnitude) 
between the constant mass approximation 
and the momentum dependent solution in 3+1 dimensions. 
The first conclusion is of course the that one cannot really trust the
constant mass approximation, as already stated above. However, such a 
change will make itself felt and, if this analysis is correct, the
relevant chiral symmetry breaking triggered by external magnetic fields
should have measurable physical consequences.

\subsection{Fermion condensate}

The fermion condensate in the LLL approximation is given by:

\be
\left<0|\ol\psi\psi|0\right>=
\frac{|eB|^{3/2}}{4\pi^2}\int_0^\infty d\rho
\frac{\mu_\rho}{Z_\rho^2\rho+\mu_\rho^2}
\ee

\nin Since we compute the condensate with the momentum dependent mass and
wave function renormalization, we do not have to compensate any 
divergence: the condensate is convergent. In a perturbative expansion,
the condensate has to be renormalized to any order but here we 
started from the  Schwinger-Dyson equation which is non-perturbative
and leads us to a convergent resummation of graphs for the condensate.
We can note that the divergence in the computation
of the fermion condensate for the constant mass 
approximation can be removed if we substract the condensate for $|eB|=0$.
We obtain then a convergent expression which 
vanishes as $m^2\to 0$. To see this, let us go back to 
the Schwinger representation (\ref{schwingerrep}) of the fermion propagator
and take away the gauge dynamics, putting $\alpha\to 0$.
The computation of the condensate is straightforward \cite{fkm} and leads to:

\be
\left<0|\ol\psi\psi|0\right>-\left<0|\ol\psi\psi|0\right>_{|eB|=0}=
\frac{m|eB|}{4\pi^2}\int_0^\infty \frac{ds}{s}e^{-s}
\left[\coth\left(s\frac{|eB|}{m^2}\right)-\frac{m^2}{s|eB|}\right]
\ee

\begin{figure}
\epsfxsize=10cm
\epsfysize=8cm
\centerline{\epsfbox{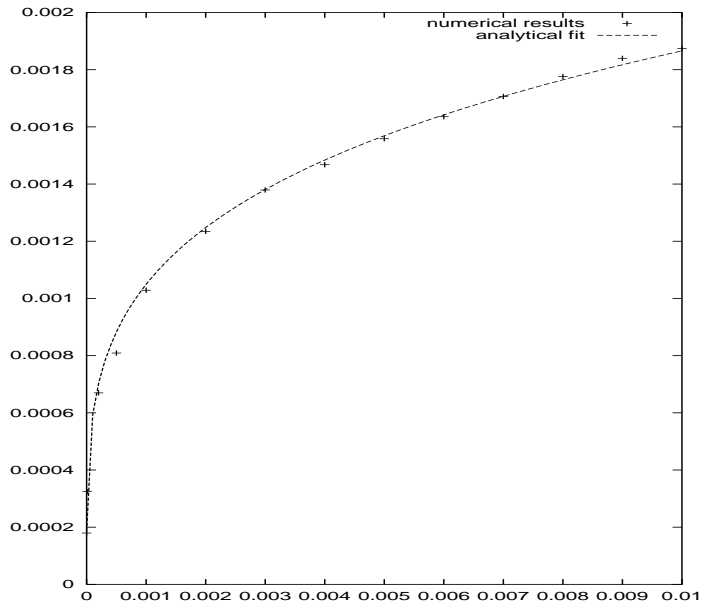}}
\caption{Dimensionless fermion condensate $B^{-3/2}<0|\ol\psi\psi|0>$
versus $\alpha$ (d=3+1)}
\label{cond4}
\end{figure}

\nin  In our numerical study with zero bare mass, we also find
a vanishing condensate for $\alpha\to 0$,
at least for strong magnetic fields
since we only considered the LLL approximation. Thus in this approximation
we do not find any critical coupling for the condensate to be generated.  
We plot in figure \ref{cond4} the dimensionless fermion condensate 
$c_f=B^{-3/2}\left<0|\ol\psi\psi|0\right>$ 
versus the coupling $\alpha$,
taking into account the relation $e^2=4\pi\alpha$. 

We note that the analytical approximation of the fermion condensate
given in \cite{lee} does not fit to our numerical results. The difference
lies in their taking the constant mass approximation and a cut-off
$|eB|$ in the momentum space. Instead we found the analytical fit:

\be
c_f^{fit}=c_0~ \alpha^{1/4}~~~~~\mbox{with}
~~~~~c_0=.00590 \pm 0.00001
\ee

\nin for the dimensionless fermion condensate.

\section{Conclusions}

Our results for the (2+1)-dimensional model are relevant for
condensed matter Physics, in particular models for high-$T_c$
superconductivity, while the (3+1)-dimensional case has to do
with the Electroweak Phase Transition in the Early Universe.

\begin{itemize}

\item As we already stated in the beginning, there has been 
a suggestion \cite{shankar, farakmavro} that 
the high-$T_c$ superconducting materials 
can be described by an effective theory, namely  a relativistic 
gauge field theory in 2+1 dimensions. The complete model was 
$SU(2) \times U(1)$ symmetric with a doublet of massless fermions \cite{fm}.  
However, simpler versions, such as QED in 2+1 dimensions have been
studied in the same context. The theory describes the quasi-particle 
excitations about the nodes of a d-wave superconducting gap. The
Fermi velocity coincides with the effective velocity of light.
In a recent experiment \cite{krishana} it has been found that for strong
magnetic fields (about O(1) to O(10) Tesla) there appear plateaus in the 
thermal conductivity of the material, signalling the opening of a new gap 
at the nodes, induced by the magnetic field. The critical temperature 
$T_{crit}$ for the appearance of this superconducting gap scales with 
the magnetic field as $T_{crit} \approx \sqrt{|e B|}.$ Assuming that
the critical temperature (where the dynamical mass vanishes) 
is roughly proportional to the dynamical mass at zero temperature, we
see that our equation (\ref{exper}), 
which says that the dynamical mass scales as
$\sqrt{|e B|},$ agrees with this experimental result. Of course,
a true finite temperature calculation must be done if a better founded
assertion can be made.

\item Concerning the dynamical mass generation and the fermionic 
condensate in 3+1 dimensions, we mention that around the Electroweak era 
magnetic fields of the order of $10^{23} - 10^{24}$ are naturally obtained 
\cite{early}. These correspond to a dynamical mass of the order of 1GeV, 
if we set $\frac{e^2}{4 \pi} \simeq \frac{1}{137},$ independently from
the vacuum expectation value of the scalar field. This is already a big 
value for the mass with eventually observable contribution in the 
effective potential at high temperature. However, if we want to give 
an explanation for the galactic magnetic fields (of the order of 
$10^{-6}$ Gauss), the estimate for the magnetic fileds during the
Electroweak Phase Transition raises to $10^{32} - 10^{33}$ Gauss, 
according to \cite{ambjorn}. If this is the case, the dynamical mass is of 
the order of 100 GeV and we believe that this possibility deserves 
further study in connection with Cosmology and the Electroweak Phase 
Transition.

\end{itemize}

\section*{Acknowledgements}

This work has been within the 
TMR project ``Finite temperature phase transitions in particle Physics",
EU contract number: FMRX-CT97-0122.
The authors would like to acknowledge financial support from the 
TMR project. Illuminating discussions with G.Triantafillou are 
gratefully acknowledged. 

\begin{appendix}

\section{Polarization tensor in magnetic field (d=2+1)}

The free photon propagator in the presence of a magnetic field 
is \cite{shpagin}

\be
D_{\mu\nu}(q)=\frac{1}{q^2}\left(g_{\mu\nu}^\|-
\frac{q_\mu^\bot q_\nu^\bot}{q^2_\bot}-\frac{q_\mu q_\nu}{q^2}\right)
+\frac{1}{q^2}\left(g_{\mu\nu}^\bot+\frac{q_\mu^\bot q_\nu^\bot}{q^2_\bot} 
\right)+\lambda\frac{q_\mu q_\nu}{q^4}
\ee

\nin where $\lambda$ is the gauge fixing parameter.
Because of the projection operator $(1-i\gamma^1\gamma^2sg(eB))/2$
that we have in the fermion propagators (\ref{freeprop}) and
(\ref{fullprop}), the components of $D_{\mu\nu}$ which include only the
transverse coordinates $(1,2)$ will not play any role in the Schwinger-Dyson
equation. Therefore in the LLL approximation the free photon 
propagator will be

\be\label{phot3}
D^{LLL}_{\mu\nu}(q)=
-\frac{i}{q^2}\left(g^\|_{\mu\nu}-\frac{q_\mu q_\nu}{q^2}\right)
-i\lambda\frac{q_\mu q_\nu}{q^4}
\ee

\nin We show now that the corrections to the
photon propagator in dimension 2+1 need   
not be taken into account in the LLL approximation.
In Euclidean space-time the one-loop level polarization tensor 
is given by \cite{farakout,tsai}

\be
\Pi_{\mu\nu}(p)=(p^2\delta_{\mu\nu}-p_\mu p_\nu)N_0(p)+
(p_\bot^2\delta_{\mu\nu}^\bot-p_\mu^\bot p_\nu^\bot)N_1(p)
\ee

\nin where

\bea
N_0(p)&=&-\frac{\alpha}{2\sqrt\pi}\int_0^\infty\frac{ds}{s}\int_{-1}^1dv
e^{-s\phi_0}\frac{z}{\sinh(z)}[\cosh(zv)-v\coth(z)\sinh(zv)]\nonu
N_1(p)&=&-\frac{\alpha}{2\sqrt\pi}\int_0^\infty\frac{ds}{s}\int_{-1}^1dv
e^{-s\phi_0}\frac{2z}{\sinh^3(z)}[\cosh(z)-\cosh(zv)]-N_0(p)
\eea

\nin with $z=|eB|s$ and 

\be
\phi_0=m^2+\frac{1-v^2}{4}p_0^2+
\frac{\cosh(z)-\cosh(zv)}{2z\sinh(z)}p_\bot^2
\ee 

\nin We will need only the component $\Pi_{00}(p)$ in the LLL
approximation, thus we only need to look at $N_0(p)$ in the 
strong field limit $|eB|>>s^{-1}$. Making an expansion for $z>>1$
we obtain after the interation over $v:$

\be
N_0(p)=-2\frac{\alpha}{\sqrt\pi}\int_0^\infty\frac{ds}{s}
z e^{-s(m^2+p_0^2/4)-z}\left(1+{\cal O}(z^{-1})\right)
\ee

\nin which finally leads to:

\be
N_0(p)=-\frac{2\alpha}{\sqrt{\pi |eB|}}\int_0^\infty d\sigma\sqrt\sigma
e^{-\sigma}\left[1+{\cal O}\left(\frac{p_0^2+m^2}{|eB|}\right)\right]
\ee

\nin We see then that the one-loop polarization
tensor vanishes as $\alpha|eB|^{-1/2}$.

\end{appendix}

\end{document}